\documentclass[aps,superscriptaddress]{revtex4}

\usepackage{graphicx}  
\usepackage{amssymb}
\usepackage{amsmath}
\usepackage{bm}

\newcommand{\Tr}{ {\rm Tr} \, }
\newcommand{\be}{\begin{equation}}
\newcommand{\ee}{\end{equation}}
\newcommand{\bea}{\begin{eqnarray}}
\newcommand{\eea}{\end{eqnarray}}
\newcommand{\hq}{\hat \psi}
\newcommand{\bS}{\mathbf{\Sigma}}
\newcommand{\bSS}{\mathbf{S}}
\newcommand{\bP}{\vec{P}}
\newcommand{\bL}{\vec{L}}
\newcommand{\bG}{\mathbf{G}}

\newcommand{\bh}{\mathbf{h}}
\newcommand{\bI}{\mathbf{I}}
\newcommand{\bg}{\mathbf{g}}
\newcommand{\bff}{\vec{f}}
\newcommand{\bE}{\vec{E}}
\newcommand{\bcalE}{\mbox{\boldmath ${\cal E}$}}
\newcommand{\bcalV}{\mbox{\boldmath ${\cal V}$}}
\newcommand{\bA}{\vec{A}}
\newcommand{\bB}{\vec{B}}
\newcommand{\bj}{\vec{j}}
\newcommand{\br}{\vec{r}}
\newcommand{\bz}{\bar{z}}
\newcommand{\bt}{\bar{t}}    
\newcommand{\vf}{\varphi}
\newcommand{\ve}{\varepsilon}
\newcommand{\Mtz}{\mathrm{M}}
\newcommand{\ret}{\mathrm{R}}
\newcommand{\adv}{\mathrm{A}}
\newcommand{\KS}{\mathrm{s}}
\newcommand{\xc}{\mathrm{xc}}
\newcommand{\x}{\mathrm{x}}
\newcommand{\Hxc}{\mathrm{Hxc}}
\newcommand{\Hartree}{\mathrm{H}}
\newcommand{\D}{\mathrm{d}}

\begin{document}

\title{Introduction to the Keldysh Formalism and Applications to 
Time-Dependent Density-Functional theory}

\author{Robert van Leeuwen}
\author{Nils Erik Dahlen}
\affiliation{Theoretical Chemistry, Materials Science Centre, Rijksuniversiteit Groningen, Nijenborgh 4, 9747 AG Groningen, The Netherlands}
\author{Gianluca Stefanucci}
\author{Carl-Olof Almbladh}
\author{Ulf von Barth}
\affiliation{Department of Solid State Theory, Institute of Physics, Lund University, S\"olvegatan 14 A, 223 62 Lund, Sweden}

\begin{abstract}

This paper gives an introduction to the Keldysh formalism, with emphasis on
its usefulness in time-dependent density functional theory. In the first part
we introduce the Keldysh contour and the one-particle Green function
defined on this contour. We then discuss how to combine  and manipulate
functions with time-arguments on the contour. The effects of electron-electron
interaction can be taken systematically into account, as we illustrate by
propagating the Kadanoff-Baym equations for the second-order self-energy
approximation. It is important to use conserving
approximations such that the evolution of the electron density, momentum, and
total energy agrees with the macroscopic conservation laws. One of the main 
topics in this paper is the non-interacting Green function, which is the 
relevant quantity for time-dependent density functional theory. We discuss
this Green function in detail, and show how the Keldysh contour
in a simple way allows us to derive the time-dependent Kohn-Sham potential
from an action functional. The formalism in a similar way leads to response
functions that obey the causality principle.
To illustrate these points, we discuss the 
time-dependent optimized effective potential equations.

\end{abstract}

\maketitle 

\section{Introduction}

We will in this paper give an introduction to the Keldysh formalism, which
is an extremely useful tool for first-principles studies of nonequilibrium 
many-particle systems. Of particular interest for TDDFT is the relation to
non-equilibrium Green functions (NEGF), which allows us to construct
exchange-correlation potentials with memory by using diagrammatic techniques.
For many problems, such as, e.g., quantum transport or atoms in intense laser
pulses, one needs exchange-correlation functionals with memory, and Green
function techniques offer a systematic method for developing these.
The Keldysh formalism is also necessary for defining response functions in
TDDFT and for defining an action functional needed for deriving TDDFT from a
variational principle. We will in this section give an introduction to the
nonequilibrium Green function  formalism, intended to illustrate the usefulness 
of the theory. The formalism does not differ much from ordinary equilibrium
theory, the main difference being that all time-dependent functions are
definied for time-arguments on a contour, known as the Keldysh contour.

The Green function, $G(\br, t;\br',t')$ is a function of two space- and
time-coordinates, and is obviously more complicated  than the one-particle
density $n(\br,t)$, which is the main ingredient in TDDFT. However, the
advantage of the 
NEGF methods is that we can systematically
improve the approximations by taking into account particular physical processes 
(represented in the form of Feynman diagrams) that we believe to be important.
The Green function provides us directly with all expectation values
of one-body operators (such as the density and the current), and also 
the total energy, ionization potentials, response functions, spectral
functions, etc.. In relation to TDDFT, this is useful not only for developing
orbital functionals and exchange-correlation functionals with memory, but also
for providing insight in the exact properties of the non-interacting Kohn-Sham
system. 

In the following, we shall focus on systems that are initially in thermal
equilibrium.  We will start by introducing the Keldysh contour and
the nonequilbrium Green functions, and then explain how to
combine and manipulate functions with time variables on the contour.
While we in TDDFT
take exchange- and correlation-effects 
into account through $v_\xc[n]$, the corresponding quantity in Green function
theory is the self-energy $\Sigma[G]$. Just like $v_\xc$, the self-energy
functional must be approximated. For a given functional $\Sigma[G]$,
it is important that the resulting observables obey the macroscopic 
conservation laws, such as, e.g., the continuity equation. These approximations
are known as \textit{conserving}, and will be discussed briefly.
In the last part of this section we will discuss the applications
of the Keldysh formalism in TDDFT, including the relation between $\Sigma$
and $v_\xc$, the derivation of the Kohn-Sham equations from an action 
functional, and the derivation of an $f_\xc$ functional. As an illustrative
example, we will discuss the time-dependent exchange-only optimized effective
potential approximation.

\section{The Keldysh Contour}

\begin{figure}
\begin{center}
\includegraphics[width=.5\textwidth]{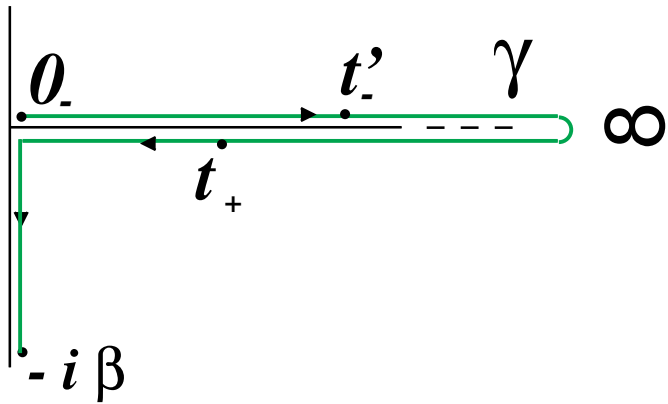}
\end{center}
\caption{The Keldysh contour, starting at $t=0$, and ending at $t=-i \beta$,
with $t$ on the backward branch and $t'$ on the forward branch.
By definition, any point lying on the vertical track is later than a 
point lying on the forward or backward branch.}
\label{ke}
\end{figure}

In quantum mechanics we associate with any observable quantity 
$O$ a hermitean operator $\hat{O}$. The expectation value $\Tr\{\hat \rho_0
\hat O\}$ gives the value of $O$ when the system is described by the density 
operator $\hat \rho_0$ and the trace denotes a sum over a complete set of states
in Hilbert space. For an isolated system 
the Hamiltonian $\hat{H}_{0}$ does not depend on time, and the expectation 
value of \textit{any} observable quantity is constant, provided $[\hat \rho_0, 
\hat H_0]=0$.  In these notes we want to discuss how to 
describe systems that are isolated for times $t<0$, such that 
$\hat H (t<0)=\hat H_0$, but disturbed by an external time-dependent
field at $t>0$. The expectation
value of $\hat O$ at $t>0$ is then given by the average on the initial 
density operator $\hat \rho_0$ of the operator $\hat{O}$ in the Heisenberg 
representation,
\begin{equation}
O (t) = \langle \hat O_H(t) \rangle \equiv \Tr\{\hat \rho_0 \hat O_H(t) \}
=\Tr\{ \hat \rho_0 \hat S(0;t) \hat O \hat S(t;0) \} 
\label{evba},
\end{equation}
where the operator in the Heisenberg picture has a time-dependence 
according to $\hat O_H(t)=\hat S(0;t) \hat O \hat S(t;0)$. The
evolution operator $\hat{S}(t;t')$ is the solution of the equations
\begin{equation}
i\frac{\D}{\D t}\hat S(t;t') = \hat H(t) \hat S(t;t') \qquad \mbox{and}
\qquad
i \frac{\D}{\D t'} \hat S(t;t') =  -\hat S(t;t')\hat H(t'),
\end{equation}
 with the boundary condition $\hat S(t;t)=1$. It
can be formally written as 
\begin{equation}
\hat{S}(t;t')=\left\{
\begin{array}{ll}
    T\,e^{-i \int_{t'}^{t} d\bar{t}\,\hat{H}(\bar{t})} &
    \quad t>t' \\
    \overline{T}\,e^{-i\int_{t'}^{t} d\bar{t}\,\hat{H}(\bar{t})} &
    \quad t<t'
\end{array}    
\right..
\label{eo}
\end{equation}
In Eq. (\ref{eo}), $T$ is the time-ordering operator that 
rearranges the operators in chronological order with later times to 
the left; $\overline{T}$ is the anti-chronological time-ordering operator.  
The evolution operator satisfies the group property 
$\hat{S}(t;t_{1})\hat{S}(t_{1};t')=\hat{S}(t;t')$ for any $t_{1}$. Notice
that if the Hamiltonian is time-independent in the interval between $t$ and
$t'$, then the evolution operator becomes $\hat{S} (t;t')=e^{-i\hat H (t-t')}$.
If we now let the system be initially in thermal equilibrium, 
with an inverse temperature $\beta\equiv 1/k_BT$ and chemical potential $\mu$, the
initial density matrix is $\hat \rho_0 =  e^{-\beta(\hat H_0 - \mu \hat N)}
/\Tr\{ e^{-\beta(\hat H_0 - \mu \hat N)}\}$. Assuming that 
$\hat H_0$ and $\hat N$ commute, $\hat \rho_0$  
can be rewritten using the evolution operator $\hat S$ with a complex
time-argument, $t=-i\beta$, according to
$ \hat \rho_0 = e^{\beta \mu \hat N} \hat S(-i\beta;0)/\Tr\{
e^{\beta \mu \hat N} \hat S(-i \beta;0) \}$. Inserting this expression 
in (\ref{evba}), we find
\begin{equation}
O(t)=\frac{\Tr \left[e^{\beta \mu \hat{N}} \hat S(-i\beta;0) \hat S(0;t) 
\hat{O} \hat S(t;0) \right]}{\Tr\left[e^{\beta \mu\hat{N}} \hat S(-i\beta;0)
\right]}.
\label{ev7}
\end{equation}    
Reading the arguments in the numerator from the right to the  left, we see
that we can design a time-contour $\gamma$ with a forward branch going 
from $0$ to $t$, a backward branch coming back from $t$ 
and ending in $0$, and a branch along the imaginary time-axis from $0$
to $-i\beta$. This contour is illustrated in Fig.~\ref{ke}. Note that 
the group property of $\hat S$ means that we are free to extend
this contour up to infinity. We can now generalize (\ref{ev7}), and let $z$ 
be a time-contour variable on $\gamma$. Letting the variable $\bz$ 
run along this same contour, (\ref{ev7}) can be formally 
recast as
\begin{equation}
O(z)
=\frac{\Tr\left[ e^{\beta\mu\hat{N}} T_\mathrm{c}\left\{e^{-i \int_{\gamma}\D 
\bz\,\hat{H}(\bz)}
\;\hat{O}(z)\right\}\right]}{\Tr\left[e^{\beta\mu\hat{N}} T_\mathrm{c}
\left\{e^{-i\int_{\gamma}\D \bz\,\hat{H}(\bz)}\right\}\right]}.
\label{ev8}
\end{equation}   
The contour ordering operator $T_{\mathrm{c}}$ moves the operators 
with ``later'' contour variable to the left. 
In (\ref{ev8}), $\hat{O}(z)$ is {\em not} the operator in the 
Heisenberg representation [the latter is denoted with $\hat{O}_{H}(t)$]. 
The contour-time argument in $\hat{O}$ is there only to
specify the position of the operator $\hat{O}$ on $\gamma$. 
A point on the real axis can be either
on the forward (we denote these points $t_-$), or on the backward branch
(denoted $t_+$), and a point which is earlier
in real time, can therefore be later on the contour, as illustrated in
Fig.~\ref{ke}.

If $z$ lies on the vertical track, then there  is no need to extend
the contour along the real axis. Instead, we have
$$
O(z)=\frac{\Tr \left[e^{\beta\mu\hat{N}}e^{-i\int_{z}^{-i\beta}\D 
\bz\hat{H}_{0}}
\hat{O}e^{-i\int_{0}^{z}\D 
\bz\hat{H}_{0}}\right]}{\Tr\left[e^{-\beta 
(\hat{H}_{0}-\mu \hat{N})}\right]}=
\frac{\Tr \left[e^{-\beta(\hat{H}_{0}-\mu\hat{N})}
\hat{O}\right]}{\Tr\left[e^{-\beta
(\hat{H}_{0}-\mu \hat{N})}\right]},
$$
where the cyclic property of the trace has been used. The right hand side 
is independent of $z$ and coincides with the thermal average
$\Tr\{ \hat \rho_0 \hat O\}$. It is easy to verify that (\ref{ev8}) would give
exactly the same result for $O(t)$, where $t$ is real, if the Hamiltonian
was time-independent, i.e. $\hat H(t) = \hat H_0$ also for $t>0$.

To summarize, in (\ref{ev8}) the variable $z$ lies on the contour of
Fig.~\ref{ke}; the r.h.s. gives the time-dependent statistical average of the
observable $O$ when $z$ lies on the forward or backward branch, and the
statistical average before the system is disturbed when $z$ lies on the
vertical track.  

\section{Nonequilibrium Green Functions}
\label{neg}

We now introduce the nonequilibrium Green function (NEGF), which is a
function of two contour time-variables.
In order to keep the notation as light as possible, we here
discard the spin degree of freedom; the spin index may be restored 
later as needed. The field operators $\hq(\br)$, $\hq^{\dag}(\br)$ 
destroy and create an electron in $\br$ and obey the anticommutation 
relations $\{\hq(\br),\hq^{\dag}(\br')\}=\delta(\br-\br')$. We write the 
Hamiltonian $\hat{H}(t)$ as the sum of a quadratic term
\be
\hat{h}(t)=\int \D\br\D\br'\;\hq^{\dag}(\br)\langle\br|\bh(t)|\br'\rangle
\hq(\br') \label{eq:h}
\ee
and the interaction operator 
\be
\hat{H}_{w}=\frac{1}{2}
\int \D\br\D\br' \;\hq^{\dag}(\br)\hq^{\dag}(\br')w(\br,\br')\hq(\br')
\hq(\br).
\ee
We use boldface to indicate matrices in one-electron 
labels, e.g., $\bh$ is a matrix and 
$\langle\br|\bh|\br'\rangle$ is the $(\br,\br')$ matrix element of $\bh$. When
describing electrons in an electro-magnetic field, the quadratic term is given
by $\langle \br | \bh(t) | \br' \rangle=\delta(\br-\br')
\left\{[\nabla/i+\bA(\br,t)]^2/2+v(\br,t)\right\}$.

The definition of an expectation value in (\ref{evba}) can be generalized to
the expectation value of two operators. The Green function is defined as
\begin{equation}
G(\br,z;\br',z')=\langle\br|\bG(z;z')|\br'\rangle\equiv -i
\langle T_\mathrm{c} [ \hq_H(\br,z)\hq_H^{\dag}(\br',z') ] \rangle,
\label{eq:gdef}
\end{equation}
where the contour variable in the field operators specifies the position 
in the contour ordering. The operators have a time-dependence according to
the definition of the Heisenberg picture, e.g. $\hq_H^{\dag}(\br,z)
=\hat S(0;z) \hq^{\dag}(\br) \hat S(z;0)$. Notice that if the time-argument
$z$ is located on the real axis, then $\hq_H(\br,t_+)=\hq_H(\br,t_-)$. If the
time-argument is on the imaginary axis, then $\hq(\br, -i\tau)$ is \textit{not}
the adjoint of $\hq(\br,-i\tau)$ since $\hat S^\dagger(-i\tau;0)\neq \hat 
S(0;-i\tau)$. The Green function can  be written
\be
\bG(z;z')=\theta(z,z')\bG^{>}(z;z')+\theta(z',z)\bG^{<}(z;z').
\label{kgf}
\ee
The function $\theta(z,z')$ is defined to be 1 if $z$ is later on the contour
than $z'$, and 0 otherwise. From the definition of the time-dependent 
expectation value in Eq.~(\ref{ev7}), it follows that the greater Green
function $\bG^{>}(z;z')$, where $z$ is later on the contour than $z'$, is
\be
G^>(\br, z;\br',z')
= \frac{1}{i}\frac{\Tr\left[e^{\beta\mu\hat{N}} \hat S(-i\beta;0)
\hq_{H}(\br,z)\hq^{\dag}_{H}(\br',z')\right]}
{\Tr\left[e^{\beta\mu\hat{N}}\hat S(-i\beta;0)\right]}.
\label{eq:ggt}
\ee
If $z'$ is later on the contour than $z$, then the Green function equals
\begin{equation}
G^<(\br, z;\br',z')=
-\frac{1}{i}\frac{\Tr\left[e^{\beta\mu\hat{N}}\hat S(-i\beta;0)
\hq^{\dag}_{H}(\br',z')\hq_{H}(\br,z)\right]}
{\Tr\left[e^{\beta\mu\hat{N}} \hat S(-i\beta;0) \right]}. \label{eq:gls}
\end{equation}
The extra minus sign on the right hand side comes from the contour ordering.
More generally, rearranging the field operators $\hq$ and $\hq^{\dag}$ (later 
arguments to the left), we also have to multiply by $(-1)^{P}$, 
where $P$ is the parity of the permutation.
From the definition of the Green function, it is easily seen that the electron
density, $n(\br,z)=\langle \hq_H^{\dag}(\br,z)\hq_H(\br,z) \rangle$ and current
is obtained according to 
\be
n(\br, z) = -i G(\br,z;\br,z^+)
\label{eq:dens}
\ee
\be
\bj (\br, z) = -i \left\{ \left[ 
\frac{\nabla}{2i} -\frac{\nabla'}{2i}  + \bA (\br,z) \right] G(\br, z;
\br',z') \right\}_{z'=z^+} .
\label{eq:curr}
\ee
where $z^+$ indicates that this time-argument is infinitesimally later on the
contour.

The Green function $\bG(z;z')$ obeys an important cyclic relation 
on the Keldysh contour. Choosing 
$z=0_{-}$, which is the earliest time on the contour, we find
$\bG(0_-;z')=\bG^<(0;z')$, given by (\ref{eq:gls}) 
with $\hq_{H}(\br,0)=\hq(\br)$.
Inside the trace we can move $\hq(\br)$ to the left. Furthermore, we 
can exchange the position of $\hq(\br)$ and $e^{\beta\mu\hat{N}}$ by 
noting that $\hq(\br)e^{\beta\mu\hat{N}}=e^{\beta\mu(\hat{N}+1)}\hq(\br)$. 
Using the group identity $\hat S(-i\beta;0) \hat S(0;-i\beta)=1$, we obtain 
\bea
G(\br,0_{-};\br',z') 
&=&
-\frac{1}{i}\frac{\Tr\left[\hq_H(\br)e^{\beta\mu\hat{N}}\hat S(-i\beta;0)
\hq_H^{\dag}(\br',z')\right]}
{\Tr\left[e^{\beta\mu\hat N } \hat S(-i\beta;0) \right]}.
\nonumber \\
&=&
-\frac{e^{\beta\mu}}{i}\frac{\Tr\left[e^{\beta\mu\hat{N}}\hat S(-i\beta;0)
\hq_H(\br,-i\beta)\hq_H^{\dag}(\br',z')\right]}
{\Tr\left[e^{\beta\mu\hat N } \hat S(-i\beta;0) \right]}.
\eea
The r.h.s. equals $-e^{\beta\mu}\langle\br|\bG(-i \beta;z')|\br'\rangle$.
Together with a similar analysis for $\bG(z;0_-)$, we conclude that
\begin{equation}
\bG(0_{-};z')=-e^{\beta\mu} \bG(-i\beta;z') \quad \mbox{and} \quad
\bG(z;0_{-}) =-e^{-\beta\mu}\bG(z;-i\beta)    .
\label{bcl}
\end{equation}
These equations constitute the so called Kubo-Martin-Schwinger 
(KMS) boundary conditions \cite{kubo,martin}. From the definition of the Green
function in (\ref{eq:gdef}), it is easily seen that the $\bG(z;z)$
has a  discontinuity in $z=z'$, 
\begin{equation}
\bG^>(z;z)=\bG^<(z;z)-i\mathbf{1}. \label{eq:disc}
\end{equation}
Furthermore, for both time-arguments on the real axis we have the important
symmetry
$\left[\bG^\lessgtr(t'; t)\right]^\dag = -\bG^\lessgtr(t; t')$. As we shall
see, these relations play a crucial role in solving the equation of motion. 

\section{The Keldysh Book-Keeping}    
\label{kbk}

The Green function belongs to a larger class of functions of two time-contour
variables that we will refer to as Keldysh space. These functions can be
written on the form
\begin{equation}
k(z;z')=\delta(z,z') k^\delta(z)+\theta(z,z')k^>(z;z')+\theta(z',z) k^<(z;z'),
\end{equation} 
where the $\delta$-function on the contour is defined as $\delta(z,z')=
\D \theta(z,z')/\D z$ 
\footnote{In general, functions containing singularity of the form
$\D^n \delta (z,z')/\D z^n$ belongs to the Keldysh space,
see \cite{daniele}}.
These functions are somewhat complicated due to the fact that each of the 
time-arguments can be located on three different brances of the contour, as
illustrated in Fig.~\ref{ke}. Below we systematically derive a set of
identities that are commonly used for dealing with such functions and will be
used extensively in the following sections. Most of the relations are well
known \cite{langreth}, 
while others, equally important \cite{wagner}, are not. Our aim is to 
provide a self-contained derivation of all of them. A table 
at the end of the Section summarizes the main results. For those who 
are not familiar with the Keldysh contour, we strongly recommend to 
scan what follows with pencil and paper. 

It is straightforward to show that if $a(z;z')$ and 
$b(z;z')$ belong to the Keldysh space, then
\begin{equation}
c(z;z')=\int_{\gamma}\D \bz \;a(z;\bz)b(\bz;z')
\label{conv}
\end{equation}
also belongs to the Keldysh space. For any $k(z;z')$ in the Keldysh 
space we define the {\em greater} and {\em lesser} 
functions on the physical time axis
$$
k^{>}(t;t')\equiv k(t_{+};t'_{-}),\quad
k^{<}(t,t')\equiv k(t_{-};t'_{+}).
$$
We also define the following two-point functions with one argument 
$t$ on the physical time axis and the other $\tau$ on the vertical track
\begin{equation}
k^{\rceil}(t;\tau)\equiv k(t_{\pm};\tau),\quad
k^{\lceil}(\tau,t)\equiv k(\tau;t_{\pm}).
\label{kceil}
\end{equation}
In the definition of $k^{\rceil}$ and $k^{\lceil}$ we can arbitrarily 
choose $t_{+}$ or $t_{-}$ since $\tau$ is later than both of them. The 
symbols ``$\rceil$'' and ``$\lceil$'' have been chosen in order to 
help the visualization of the time arguments. For instance, 
``$\rceil$'' has a horizontal segment followed by a vertical one;  
correspondingly, $k^{\rceil}$ has a first argument which is real 
(and thus lies on the horizontal axis) and a second argument which is 
imaginary (and thus lies on the vertical axis). We will also use the 
convention of denoting the real time with latin letters and the imaginary 
time with greek letters.

If we write out the contour  integral in (\ref{conv}) in detail, we see 
with the help of Fig.~\ref{ke} that the integral consists of four main parts.
First, we must integrate along the real axis from $\bz=0_-$ to $\bz=t'_-$, for
which $a=a^>$ and $b=b^<$. Then, the  integral goes from $\bz=t'_-$ to
$\bz=t_+$, where  $a=a^>$ and $b=b^>$. The third part of the integral goes
along the
real axis from $\bz=t_+$ to $\bz=0_+$, with  $a=a^<$ and $b=b^>$. The last 
integral is along the imaginary track, from $0_+$ to $-i \beta$, where
$a=a^\rceil$ and $b=b^\lceil$. In addition, we have the contribution from
the singular parts, $a^\delta$ and $b^\delta$, which is trivial since these
integrals involve a $\delta$-function. With these specifications, we can drop
the $\pm$-subscripts on the time-arguments and write
\begin{eqnarray}
c^{>}(t;t')
&=& a^>(t,t') b^\delta (t')+ a^\delta (t) b^>(t,t')+ 
\int_{0}^{t'}\D \bar{t} \;a^{>}(t;\bar{t})b^{<}(\bar{t};t') \nonumber\\
&&+ \int_{t'}^{t}\D \bar{t}\;a^{>}(t;\bar{t})b^{>}(\bar{t};t')
+
\int_{t}^{0}\D \bar{t}\;a^{<}(t;\bar{t})b^{>}(\bar{t};t')
+
\int_{0}^{-i \beta}\D \bar{\tau}\;a^{\rceil}(t;\bar{\tau})b^{\lceil}(\bar{\tau};t').
\nonumber
\end{eqnarray}
The second integral on the r.h.s. is an ordinary integral on the 
real axis of two well defined functions and may be rewritten as
$$
\int_{t'}^{t}\D\bar{t}\;a^{>}(t;\bar{t})b^{>}(\bar{t};t')=
\int_{t'}^{0}\D\bar{t}\;a^{>}(t;\bar{t})b^{>}(\bar{t};t')
+\int_{0}^{t}\D\bar{t}\;a^{>}(t;\bar{t})b^{>}(\bar{t};t').
$$
Using this relation, the expression for $c^{>}$ becomes
\begin{eqnarray}
c^{>}(t;t')&=& a^>(t,t') b^\delta(t') + a^\delta(t) b^>(t,t')
-\int_{0}^{t'}\D \bar{t} \;a^{>}(t;\bar{t})[b^{>}(\bar{t};t')-
b^{<}(\bar{t};t')]
\nonumber \\ &+&
\int_{0}^{t}\D \bar{t}\;[a^{>}(t;\bar{t})-a^{<}(t;\bar{t})]b^{>}(\bar{t};t')
 +
\int_{0}^{-i\beta}\D\bar{\tau}\;a^{\rceil}(t;\bar{\tau})b^{\lceil}(\bar{\tau};t').
\label{cgc}
\end{eqnarray}

Next, we introduce two other functions on the physical time axis 
\begin{eqnarray}
k^\ret (t;t')&\equiv& \delta(t,t') k^\delta 
+\theta(t-t')[k^{>}(t;t')-k^{<}(t;t')],
\label{kr} \\
k^\adv (t;t')&\equiv&\delta(t,t') k^\delta 
-\theta(t'-t)[k^{>}(t;t')-k^{<}(t;t')].
\label{ka}
\end{eqnarray}
The {\em retarded} function $k^\ret (t;t')$ vanishes for $t<t'$, while 
the {\em advanced} function $k^\adv (t;t')$ vanishes for $t>t'$. The 
retarded and advanced functions can be used to rewrite (\ref{cgc}) 
in a more compact form
$$
c^{>}(t;t')=\int_{0}^{\infty}\D\bar{t}\,[a^{>}(t;\bar{t})b^\adv (\bar{t};t')
+a^\ret (t;\bar{t})b^{>}(\bar{t};t')]
+\int_{0}^{-i\beta}\D\bar{\tau}\,a^{\rceil}(t;\bar{\tau})b^{\lceil}(\bar{\tau};t').
$$
It is convenient to introduce a short hand notation for integrals 
along the physical time axis and for those between 0 and $-i\beta$. The 
symbol ``$\cdot$'' will be used to write 
$\int_{0}^{\infty}\D\bt f(\bt)g(\bt)$ as $f\cdot g$, while 
the symbol ``$\star$'' will be used to write 
$\int_{0}^{-i \beta}\D\bar{\tau}f(\bar{\tau})g(\bar{\tau})$ as $f\star g$. Then
\begin{equation}
c^{>}=a^{>}\cdot b^\adv +a^\ret \cdot b^{>}+a^{\rceil}\star 
b^{\lceil}.
\label{c>}
\end{equation}
Similarly, one can prove that
\begin{equation}
c^{<}=a^{<}\cdot b^\adv+a^\ret \cdot b^{<}+a^{\rceil}\star 
b^{\lceil}.
\label{c<}
\end{equation}

Equations (\ref{c>}-\ref{c<}) can be used to extract the retarded 
and advanced component of $c$. By definition
\begin{eqnarray}
c^\ret (t;t')&=&\delta(t-t') c^\delta(t) + \theta(t-t')[c^{>}(t;t')-c^{<}(t;t')]
\nonumber \\ 
&=&a^\delta(t) b^\delta(t') \delta(t-t')+ \theta(t-t')\int_{0}^{\infty}\D\bar{t}\,
a^\ret (t;\bar{t})[b^{>}(\bar{t};t')- b^{<}(\bar{t};t')]
\nonumber \\
&&+\theta(t-t')\int_{0}^{\infty}\D\bar{t}\,
[a^{>}(t;\bar{t})-a^{<}(t;\bar{t})]b^\adv (\bar{t};t').
\nonumber
\end{eqnarray}
Using the definitions (\ref{kr}) and (\ref{ka}) to expand the integrals on the
r.h.s. of this equation, it is straightforward to show that
\begin{equation}
c^\ret =a^\ret \cdot b^\ret .
\label{cr}
\end{equation}
Proceeding along the same lines, one can show that the advanced 
component is given by $c^\adv=a^\adv \cdot b^\adv $. 
It is worth noting that in the expressions for $c^\ret$ and $c^\adv$ 
no integration along the imaginary track is required.

Next, we show how to extract the components $c^{\rceil}$ and 
$c^{\lceil}$. We first define the {\em Matzubara} function 
$k^\Mtz (\tau;\tau')$ 
with both the arguments in the interval $(0,-i\beta)$:
$$
k^\Mtz (\tau;\tau')\equiv k(z=\tau;z'=\tau').
$$
Let us focus on $k^{\rceil}$. 
Without any restrictions we may take $t_{-}$ as the first argument
in (\ref{kceil}). In this case, we find
\be
c^{\rceil}(t;\tau) = a^\delta(t) b^\rceil(t;\tau) +
\int_{0_{-}}^{t_{-}}\D\bz a^{>}(t_{-};\bz)b^{<}(\bz;\tau)
 +\int_{t_{+}}^{0_{+}}\D\bz a^{<}(t_{-};\bz)b^{<}(\bz;\tau)
+ \int_{0_{+}}^{-i\beta}\;\D\bz\;a^{<}(t_{-};\bz)b(\bz;\tau).
\ee
Converting the contour integrals in integrals along the real time axis 
and along the imaginary track,
and taking into account the definition in (\ref{kr})
\begin{equation}
c^{\rceil}=a^\ret \cdot b^{\rceil}+a^{\rceil}\star b^\Mtz.
\label{crc}
\end{equation}
The relation for $c^{\lceil}$ can be obtained in a similar way and 
reads $c^{\lceil}=a^{\lceil}\cdot b^\adv+a^\Mtz \star 
b^{\lceil}$. 
Finally, it is straightforward to prove that the Matzubara component 
of $c$ is simply given by
$c^\Mtz=a^\Mtz \star b^\Mtz$. 

There is another class of identities we want to discuss for 
completeness. We have seen that the convolution (\ref{conv}) of two functions 
belonging to the Keldysh space also belongs to the Keldysh space. 
The same holds true for the product
$$
c(z;z')=a(z;z')b(z';z).
$$
Omitting the arguments of the functions, one readily finds (for $z\neq z'$)
\begin{equation}
c^{>}=a^{>}b^{<},\;\;\;
c^{<}=a^{<}b^{>},\quad
c^{\rceil}=a^{\rceil}b^{\lceil},\;\;\;
c^{\lceil}=a^{\lceil}b^{\rceil},
\quad
c^\Mtz =a^\Mtz b^\Mtz.
\label{c><p}
\end{equation}
The retarded function is then obtained exploiting the identities in
(\ref{c><p}). We have (for $t\neq t'$)
$$
c^\ret (t;t')=\theta(t-t')[a^{>}(t;t')b^{<}(t';t)-a^{<}(t;t')b^{>}(t';t)].
$$
We may get rid of the $\theta$-function by 
adding and subtracting $a^{<}b^{<}$ or $a^{>}b^{>}$ to the above 
relation and rearranging the terms. The final result is  
$$
c^\ret =a^\ret b^{<}+a^{<}b^\adv =a^\ret b^{>}+a^{>}b^\adv .
$$
Similarly one finds
$c^\adv =a^\adv b^{<}+a^{<}b^\ret =a^\adv b^{>}+a^{>}b^\ret$.
The time-ordered and anti-time-ordered functions can be obtained in a 
similar way and the reader can look at Table \ref{kid} for the complete list of 
definitions and identities.

For later purposes, we also consider the case of a Keldysh function 
$k(z;z')$ multiplied on the left by a scalar function $l(z)$. The scalar 
function is equivalent to the singular part of a function belonging to
Keldysh space, $\tilde l(z;z')=l(z)\delta(z,z')$, meaning that 
$\tilde l^{\ret/\adv} = \tilde l^\Mtz =\tilde l$ and 
$\tilde l^\lessgtr=\tilde l^\rceil=\tilde l^\lceil=0$. 
Using Table \ref{kid}, one immediately realizes that the function $l$ is simply a
prefactor: $\int_\gamma d\bar z \, \tilde l(z;\bar z) k^\x (\bar z; z')=l(z) k^\x
(z;z')$, where $\x$ 
is one of the Keldysh components ($\lessgtr$, 
$\mathrm{ R,\;A}$, $\rceil,\;\lceil$, $\mathrm{M}$).
The same 
is true for $\int_\gamma\, k^\x (z;\bar z) \tilde r(\bar z; z')=k^\x (z;z')r(z')$,
where $\tilde r(z;z')=r(z) \delta(z,z')$ and $r(z)$ is a scalar function.

\begin{table}
    \caption{Table of definitions of Keldysh functions and identities 
    for the convolution and the product of two functions in the Keldysh space.}
    \label{kid}
\begin{ruledtabular}
\begin{tabular}{@{\extracolsep\fill}ccc}
Definition  & $c(z;z')=\int_{\gamma}\D \bz \;a(z;\bz)b(\bz;z')$ & $c(z;z')=a(z;z')b(z';z)$  \\ \hline \\
	$k^{>}(t;t')=k(t_{+};t'_{-})$ & 
	$c^{>}=a^{>}\cdot b^\adv +a^\ret \cdot b^{>}+a^{\rceil}\star 
	b^{\lceil}$ &
	$c^{>}=a^{>}b^{<}$\\
	$k^{<}(t;t')=k(t_{-};t'_{+})$ & 
	$c^{<}=a^{<}\cdot b^\adv +a^\ret \cdot b^{<}+a^{\rceil}\star 
	b^{\lceil}$ &
	$c^{<}=a^{<}b^{>}$\\
	$\begin{array}{l} k^\ret (t;t')=\delta(t-t')k^\delta(t) \\ 
\hspace{1.0cm} +\theta(t-t')[k^{>}(t;t')-k^{<}(t;t')] \end{array}$ &
	$c^\ret =a^\ret \cdot b^\ret $ &
	$c^\ret =\left\{
	\begin{array}{l}a^\ret b^{<}+a^{<}b^\adv \\ 
	a^\ret b^{>}+a^{>}b^\adv \end{array}\right.$\\
	$\begin{array}{l} k^\adv(t;t')=\delta(t-t') k^\delta(t) \\
\hspace{1.0cm} -\theta(t'-t)[k^{>}(t;t')-k^{<}(t;t')] \end{array}$ &
	$c^\adv =a^\adv \cdot b^\adv $ &
	$c^\adv =\left\{\begin{array}{l}
	a^\adv b^{<}+a^{<}b^\ret \\
	a^\adv b^{>}+a^{>}b^\ret \end{array}\right.$\\
	$k^{\rceil}(t;\tau)= k(t_{\pm};\tau)$ &
	$c^{\rceil}=a^\ret \cdot b^{\rceil}+a^{\rceil}\star b^\Mtz $ &
	$c^{\rceil}=a^{\rceil}b^{\lceil}$\\
	$k^{\lceil}(\tau,t)= k(\tau;t_{\pm})$ &
	$c^{\lceil}=a^{\lceil}\cdot b^\adv +a^\Mtz \star b^{\lceil}$ &
	$c^{\lceil}=a^{\lceil}b^{\rceil}$\\
	$k^\Mtz (\tau;\tau')=k(z=\tau;z'=\tau')$ &
	$c^\Mtz =a^\Mtz \star b^\Mtz $ &
	$c^\Mtz =a^\Mtz b^\Mtz $\\
	\hline
    \end{tabular}
    \end{ruledtabular}
\end{table}

\section{The Kadanoff-Baym equations}

The Green function, as defined  in (\ref{kgf}), satisfies the equation
of motion 
\begin{equation}
i\frac{\D}{\D z} \bG(z;z')
={\bf 1}\delta(z,z')  +\bh(z)\bG(z;z') +\int_\gamma d\bar z\, \bS(z,\bar z) 
\bG(\bar z; z') , 
\label{lem}
\end{equation}
as well as the adjoint equation
\begin{equation}
-i \frac{\D}{\D z'} \bG(z;z') = {\bf 1}\delta(z,z') 
+\bG(z;z') \bh(z') +\int_\gamma d\bar z\, \bG(z;\bar z) \bS(\bar z, z').
\label{rem}
\end{equation}
The external potential is included in $\bh$, while the self-energy $\bS$ is a 
functional of the Green function, and describes the effects of the electron
interaction. The self-energy belongs to Keldysh space and can therefore be
written on the form $\bS(z,z')=\delta(z,z')
\bS^\delta(z) + \theta(z,z')\bS^>(z,z')+\theta(z',z)\bS^<(z,z')$.
The singular part of the self-energy can be identified as the Hartree--Fock
potential, $\bS^\delta(z)=\mathbf{v}_\Hartree(z)+\bS_\x(z)$.
The self-energy obeys the same anti-periodic 
boundary conditions at $z=0_-$ and $z=-i \beta$ as $\bG$. We will discuss
self-energy approximations in more detail below.

\begin{figure}
\begin{center}
\includegraphics[width=1.0\textwidth]{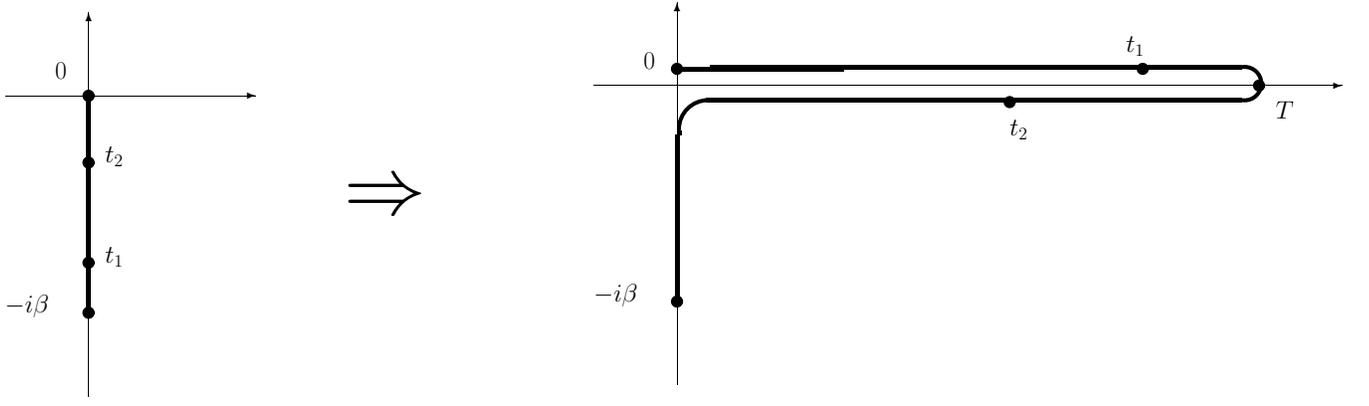}
\end{center}
\caption{Propagating the Kadanoff-Baym equations means that one first
determines the Green function for time-variables along the imaginary track.
One then calculates the Green function with one or two variables on an 
expanding time-contour.} \label{fig:contour}
\end{figure}

Calculating the Green function on the
time-contour now consists of two steps: 1) First one has to find the Green
function for imaginary times, which is equivalent to finding the equilibrium
Matzubara Green function $\bG^\Mtz (\tau,\tau')$. This Green function
depends only on the difference between the time-coordinates, and satisfies
the KMS boundary conditions according to $\bG^\Mtz (\tau+i\beta,\tau')=
-e^{\beta\mu N} \bG^\Mtz (\tau, \tau')$. Since the self-energy depends
on the Green function, this amounts to solving the finite-temperature Dyson
equation to self-consistency.
2) The Green function with one or
two  time-variables on the real axis can now be found by propagating according
to (\ref{lem}) and (\ref{rem}).
Starting from $t=0$, this procedure corresponds to 
extending the time-contour along the real time-axis. The process is
illustrated in Fig.~\ref{fig:contour}.
Writing out the equations for the 
components of $\bG$ using Table \ref{kid}, we obtain the equations known as 
the Kadanoff-Baym equations \cite{kb-book},
\be
i\frac{\D}{\D t} \bG^\lessgtr(t; t') = \bh(t) \bG^\lessgtr(t;t')
+\left[\bS^\ret \cdot \bG^\lessgtr\right] (t;t')
+\left[\bS^\lessgtr \cdot \bG^\adv \right] (t;t')
+ \left[\bS^\rceil \star \bG^\lceil \right](t,t'), \label{eq:kb1}
\ee
\be
-i\frac{\D}{\D t'} \bG^\lessgtr(t; t') = \bG^\lessgtr(t;t')\bh(t')
+\left[ \bG^\lessgtr \cdot \bS^\adv \right] (t;t')
+\left[ \bG^\ret \cdot \bS^\lessgtr \right] (t;t')
+ \left[ \bG^\rceil \star \bS^\lceil  \right](t,t'), \label{eq:kb2}
\ee
\begin{equation}
i\frac{\D}{\D t}\bG^\rceil(t;\tau) = \bh(t) \bG^\rceil(t;\tau)
+\left[\bS^\ret \cdot \bG^\rceil\right] (t; \tau)
+ \left[\bS^\rceil \star \bG^\mathrm{M} \right](t,\tau),
\label{eq:kb3}
\end{equation} 
and
\begin{equation}
-i\frac{\D}{\D t}\bG^\lceil(\tau;t) = \bG^\lceil(\tau;t) \bh(t)
+ \left[\bS^\lceil \cdot \bG^\adv \right](\tau,t)
+\left[\bS^\Mtz \star \bG^\lceil\right] (\tau; t).
\label{eq:kb4}
\end{equation} 
It is easily seen that if we denote by $T$ the largest of the two
time-arguments $t$ and $t'$, then
the right hand sides  of (\ref{eq:kb1}) and (\ref{eq:kb2}) depend on
$\bG^\lessgtr(t_1; t_2)$, $\bG^\lceil(\tau_1,t_2)$ and 
$\bG^\rceil(t_1,\tau_2)$ for $t_1, t_2
\le T$. When propagating the Kadanoff-Baym equations one therefore 
starts at $t=t'=0$, with the initial conditions given by
$\bG^<(0;0)=\lim_{\eta \to 0} \bG^\Mtz (0;-i\eta)$,
$\bG^>(0;0)=\lim_{\eta \to 0} \bG^\Mtz (-i\eta;0)$, 
$\bG^\lceil(\tau,0)=\bG^M(\tau,0)$ and $\bG^\rceil(0,\tau)=
\bG^M(0,\tau)$.  One then calculates  $\bG^\lessgtr(t,t')$ 
for time-arguments within the
expanding square given by $t,t'\le T$. Simultaneously, one calculates 
$\bG^\lceil(t,\tau)$ and $\bG^\rceil(\tau,t)$ for $t\le T$. 
The resulting $\bG$ then automatically satisfies the KMS boundary conditions.
The Kadanoff-Baym equations (\ref{eq:kb1}) and (\ref{eq:kb3}) can both be
written in the form 
\be
i\frac{\D}{\D t} \bG^\x(t;z') = \bh^{\mathrm{HF}}(t)
\bG^\x (t;z') + 
\bI^\x (t;z'),
\ee
while (\ref{eq:kb2}) and (\ref{eq:kb4}) can be written as the adjoint
equations.  The term proportional to $\bh^{\mathrm{HF}}\equiv\bh
+\bS^\delta$ describes a free-particle propagation, while $\bI^\x$ is a 
collision term, which introduces memory effects and dissipation. As can be
seen from (\ref{eq:kb1}--\ref{eq:kb4}), the only
contribution to $\bI^\x (0;0)$  comes from terms containing
time-arguments on the imaginary axis. These terms therefore contain the effect
of initial correlations, since the time-derivative of $\bG$ would otherwise
correspond to that of a non-interacting system, i.e.,
$\bI^\x (0;0)=0$.

\begin{figure}
\begin{center}
\includegraphics[width=1.0\textwidth]{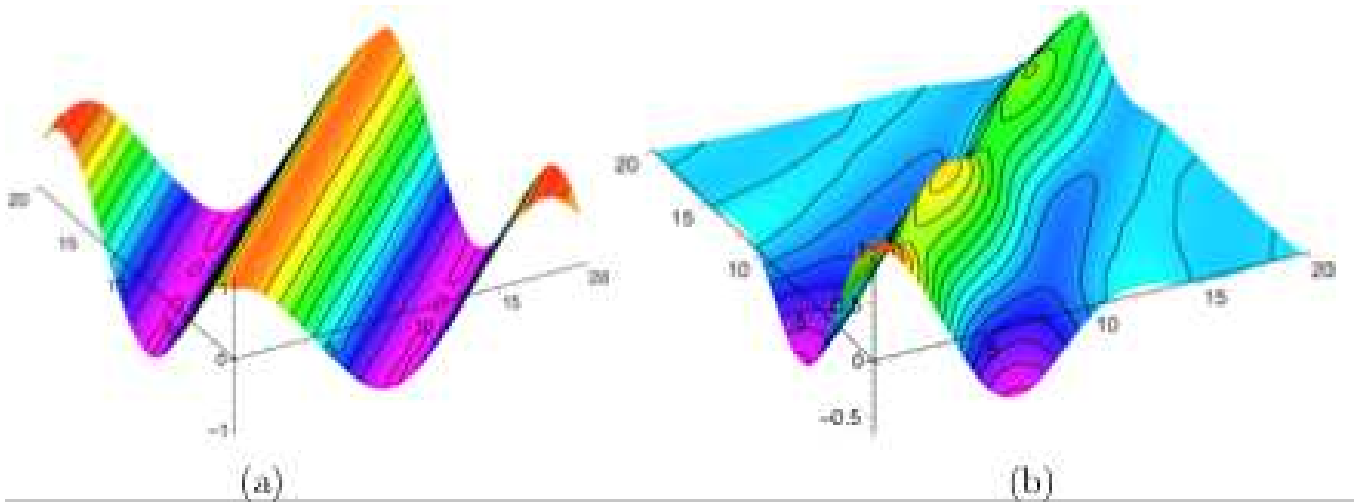}
\end{center}
\caption{The figures show the Green function
$\mathrm{Im}\, G^<_{\sigma_g, \sigma_g}(t_1, t_2)$, where
the matrix indices refer to the groundstate $\sigma_g$ Hartree--Fock orbital
of the molecule. The figure on the left shows the system in equilibrium, while
the system on the right has an additional electric field, $\theta(t) E_0 z$.
The times $t_1$ and $t_2$ on the axes are given in atomic units.}
\label{fig:dyson}
\end{figure}

An example of a time-propagation is given in
Fig.~\ref{fig:dyson}, which shows the Green function for an H$_2$ molecule.
In this example, the Green function is represented in a basis of
Hartree--Fock molecular orbitals, $\langle \br | \bG(z,z') | \br'\rangle 
=\sum_{ij} \phi_i(\br) G_{ij} (z,z')\phi^*_j(\br')$, where $\phi_i(\br)=
\langle \br | \phi_i\rangle$.
We have here propagated the Kadanoff-Baym equations using the second Born
approximation, illustrated in Fig.~\ref{fexamples}b.
The plots show the imaginary part of the matrix element 
$G^<_{\sigma_g,\sigma_g}(t,t')$ calculated for time-variables within
the square $t_1, t_2 \le T=20.0$ atomic units.
In the plot to the left, there is no added external potential and the
molecule remains in equilibrium. This means that the Green function only
depends on the difference $t_2-t_1$, and the oscillations in this 
time-coordinate has a frequency given by the ionization potential of the
molecule, in agreement with equilibrium Green function theory 
\cite{fetterwalecka}. The figure on the right shows the same matrix element, 
but now in the presence of an additional electric field which is switched on at
$t=0$. The oscillations along the ridge
$t_1=t_2$ can be interpreted as oscillations in the occupation number.

\section{Conserving approximations}

In the Dyson-Schwinger equations (\ref{lem}) and (\ref{rem}), we introduced the
electronic self-energy functional $\bS$, which accounts for the effects of the
electron interaction. The self-energy is a functional of the Green function,
and will have to be approximated in practical calculations. 
Diagrammatic techniques provide a 
natural scheme for generating approximate self-energies and for
systematically improving the approximations. There are no general 
prescriptions for how to select the relevant diagrams,
which means that this selection must be guided by physical intuition.
There are, however, important conservation 
laws, like the number conservation law or the energy conservation 
law, that should always be obeyed. We will in the following
discuss an exact framework for generating such {\em conserving 
approximations}. 

Let us first discuss the conservation laws obeyed by a system of interacting
electrons, in an external field given by the electrostatic potential
$v(\br,t)$ and vector potential $\bA(\br, t)$.
An important relation between these two quantities is provided by the
continuity equation
\be
\frac{\D}{\D t}  n(\br, t) + \nabla \cdot \bj (\br, t)  = 0.
\label{eq:number}
\ee
The density and the current density can be calculated from the Green function
using (\ref{eq:dens}) and (\ref{eq:curr}). Whether these quantities will
agree with the  continuity equation will depend on whether the Green function
is obtained from a conserving self-energy approximation.
If we know the current density
we can also calculate the total momentum and angular momentum 
expectation values in the system from the equations
\be
\bP (t) =  \int d\br \,  \bj (\br, t)  \\
\qquad \mbox{and} \qquad
\bL (t) = \int d\br \, \br \times \bj (\br, t)  .
 \ee
For these two quantities the following relations should be satisfied
\bea
\frac{\D}{\D t} \bP (t) &=& -\int d\br \Big[
n(\br, t) \, \bE (\br, t) +  \bj (\br, t) \times \bB (\br, t) \Big]
\label{eq:force} \\
\frac{\D}{\D t} \bL (t)  &=&
-\int d\br \Big[ n(\br, t)  \br \times \bE (\br, t) + 
\br \times (  \bj (\br, t) \times \bB (\br, t)) \Big] .
\label{eq:torque}
\eea
where $\bE$ and $\bB$ are the electric and magnetic fields calculated from
\be
\bE (\br, t) = \nabla v(\br,t) - \partial_{t} \bA (\br, t) 
\qquad \mbox{and} \qquad
\bB (\br, t) = \nabla \times \bA (\br, t) .
\ee
The equations (\ref{eq:force}) and (\ref{eq:torque}) tell us that the
change in momentum and angular momentum is equal to the total force and
total torque on the system. In the absence of external fields these
equations express momentum and angular momentum conservation.
Since the right hand sides of (\ref{eq:force}) and (\ref{eq:torque})
can also directly be calculated from the density and the current and therefore
from the Green function, we may wonder whether they are satisfied 
for a given approximation to the Green function.

Finally we will consider the case of energy conservation.  
Let $E (t) = \langle \hat H (t) \rangle$ be the energy expectation value of
the system, then we have
\be
\frac{\D }{\D t} E (t) = - \int d\br \, \bj (\br, t)  \cdot \bE (\br, t)
\label{eq:energy}
\ee
This equation tells us that the energy change of the system is equal to 
the work done on the system.
The total energy is calculated from the
Green function using the expression
\be
E (t) = -\frac{i}{2} \int d\br \left. \langle \br |
\left[ i \frac{\D }{\D t} + \bh(t) \right] \bG^< (t, t') | \br\rangle
\right|_{t'=t} .
\ee 
The question is now whether the energy and the current calculated from an
approximate Green function satisfy the relation in (\ref{eq:energy}).

Baym and Kadanoff \cite{baym1,baym2} showed that  conserving
approximations follow immediately if the self-energy is obtained as the 
functional derivative,
\begin{equation}
\Sigma(1;2)=\frac{\delta \Phi}{\delta G(2;1)}.
\label{fse}
\end{equation}
Here, and in the following discussion, we use numbers to denote the contour
coordinates, such that $1=(\br_1, z_1)$.
A functional $\Phi$ can be constructed, as first shown in a
seminal paper by Luttinger and Ward \cite{lw}, by summing over irreducible
self-energy diagrams closed with an additional Green function line and 
multiplied by appropriate numerical prefactors,
\begin{equation}
\Phi[G]=\sum_{n,k} \frac{1}{2n}\int d\bar 1 d\bar 2 \,
\Sigma_{n}^{(k)}(\bar{1};\bar{2})G(\bar{2};\bar{1}).
\label{exactf}
\end{equation}
In this summation, $\Sigma_{n}^{(k)}$ denotes a self-energy diagram of $n$-th
order, i.e. containing $n$ interaction lines. The time-integrals go along the
contour, but the rules for constructing Feynman diagrams are otherwise exactly
the same as those in the ground-state formalism \cite{fetterwalecka}.
Notice that the functional derivative in (\ref{fse}) may generate other
self-energy diagrams in addition to those used in the construction
of $\Phi$ in (\ref{exactf}).
\begin{figure}
\begin{center}    
\includegraphics[width=1.0\textwidth]{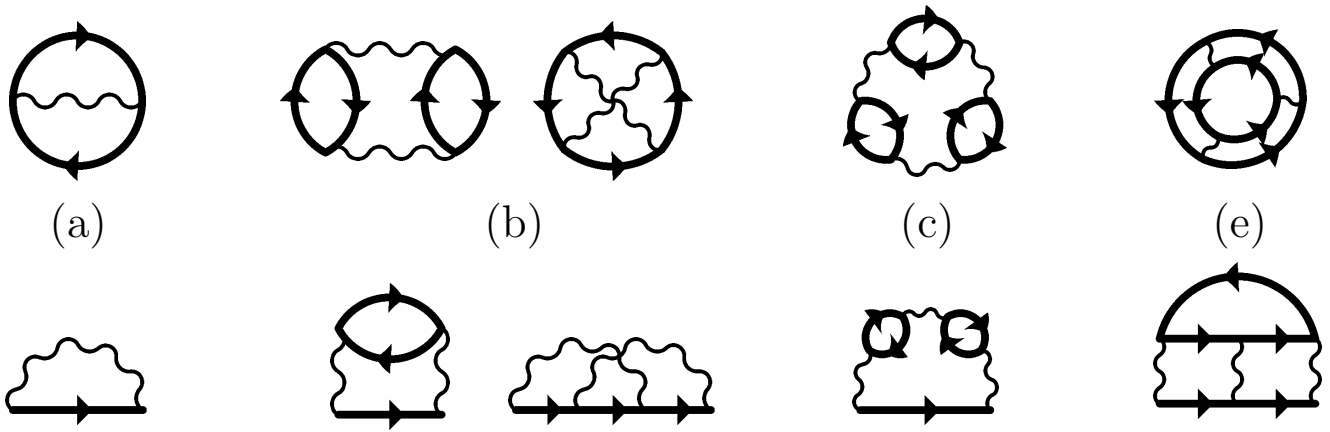}
    \caption{Diagrams for the generating functional $\Phi$, and the
    corresponding self-energy diagrams. In (a) we have the exchange
    diagram, and (b) the second Born approximation. The diagrams in
    (c) and (d) belong to the $GW$ approximation and the $T$-matrix
    approximation respectively.}
\label{fexamples}
\end{center}
\end{figure}
In Fig.~\ref{fexamples} we show some examples of typical $\Phi$ diagrams.
Examples of $\Phi$-derivable approximations include Hartree--Fock, the second
Born approximation, the $GW$ approximation and the $T$-matrix approximation.

When the Green function is calculated from a conserving approximation, the
resulting observables agree  with the conservation laws of the underlying
Hamiltonian, as given in (\ref{eq:number}), (\ref{eq:force}), 
(\ref{eq:torque}), and (\ref{eq:energy}).
This guarantees the conservation of particles, 
energy, momentum, and angular momentum. All these conservation 
laws follow from the invariance of $\Phi$ under specific changes in $G$.
We will here only outline the principles of the proofs, without going into the
details,  which can be found in \cite{baym1,baym2}.
1) \textit{Number conservation}
follows from the gauge invariance of $\Phi$. A gauge
transformation $\bA(1) \to \bA(1)+\nabla \Lambda(1)$, where
$\Lambda(\br,0_-)=\Lambda(\br,-i\beta)$ leaves $\Phi$ unchanged. 
A consequence of the gauge invariance is that a pure gauge cannot induce 
a change in the density or current. The invariance is therefore closely  
related to the Ward-identities and to the $f$-sum rule for the density response
function \cite{vanleeuwen04}. 
2) \textit{Momentum conservation} follows from the invariance of $\Phi$  under 
spatial
translations, $\br\to\br+\vec{R}(z)$. The invariance  is a consequence of
the electron interaction $v(1,2)=\delta(z_1,z_2)/|\br_1-\br_2|$ being 
instantaneous and only depending on the difference between the spatial 
coordinates.
3) \textit{Angular momentum conservation} follows from the invariance of $\Phi$
under a rotation of the spatial coordinates. 4)  \textit{Energy conservation}
follows from the invariance of $\Phi$ when described by an observer using
a ''rubbery clock'', measuring time according to the function  $s(z)$.
The invariance relies on the electron interaction being instantaneous.

\section{Non-interacting Electrons}

In this Section we focus on non-interacting electrons. This is particularly
relevant for TDDFT, where the electrons are described by the non-interacting
Kohn-Sham system. While the Kohn-Sham Green function differs from the true
Green function, they both produce the same time-dependent density. This is 
important since the density is not only an important observable in, e.g.,
quantum transport, but also since the density is the central ingredient in 
TDDFT. The use of NEGFs in TDDFT is therefore important due to the relation between
$v_\xc$ and the self-energy. 

For a system of non-interacting electrons $\hat{H}_{v}=0$ and it is 
straightforward to show that the 
Green function obeys the equations of motion (\ref{lem}) and (\ref{rem}), with
$\bS=0$.
For any $z\neq z'$, the equations of motion can be solved by using the 
evolution operator on the contour,
$$
\bSS(z,z')=T_\mathrm{c}\left\{e^{-i\int_{z'}^{z}\D\bz \;\bh(\bz)}\right\},
$$
which solves $i\frac{\D}{\D z}\bSS(z,z')=\bh(z)\bSS(z,z')$ 
and $-i\frac{\D}{\D z'}\bSS(z,z')=\bSS(z,z')\bh(z')$.
Therefore, any Green function
\be
\bG(z;z')=\theta(z,z')\bSS(z,0_{-})\bff^{>}\bSS(0_{-},z')
+\theta(z',z)\bSS(z,0_{-})\bff^{<}\bSS(0_{-},z'), \label{eq:n-g}
\ee
satisfying the constraint (\ref{eq:disc}) on the form
\begin{equation}
\bff^{>}-\bff^{<}=-i{\bf 1},
\label{cons}
\end{equation}
is a solution of the (\ref{lem}-\ref{rem}).
In order to fix the matrix $\bff^{>}$ 
or $\bff^{<}$ we impose the KMS boundary conditions. The matrix 
$\bh(z)=\bh_{0}$ for any $z$ on the vertical track, meaning 
that $\bSS(-i\beta,0_{-})=e^{-\beta\bh_{0}}$. Equation (\ref{bcl}) then 
implies $\bff^{<}=-e^{-\beta(\bh_{0}-\mu)}\bff^{>}$, and taking into 
account the constraint (\ref{cons}) we conclude that 
$$
\bff^{<}=\frac{i}{e^{\beta(\bh_{0}-\mu)}+1}=i f(\bh_{0}),
$$
where $f(\omega)=1/[e^{\beta(\omega-\mu)}+1]$ is the Fermi distribution function. 
The matrix $\bff^{>}$ takes the form $\bff^{>}=i[f(\bh_{0})-{\bf 1}]$.

The Green function $\bG(z;z')$ for a system of non-interacting electrons is now
completely fixed. Both $\bG^{>}$ and $\bG^{<}$ depend on the initial 
distribution function $f(\bh_{0})$, as it should according to the discussion
of Section \ref{neg}. Another way of writing $-i\bG^{<}$ is in terms of the
eigenstates $|\vf_n\rangle \equiv
|\vf_{n}(0)\rangle$ 
of $\bh_{0}$ with eigenvalues $\ve_{n}$. From the time-evolved eigenstate 
$|\vf_{n}(t)\rangle=\bSS(t,0)|\vf_n\rangle$ 
we can calculate the time-dependent wavefunction 
$\vf_{n}(\br,t)=\langle\br|\vf_{n}(t)\rangle$. Inserting 
$\sum_{n}|\vf_{n}(0)\rangle\langle\vf_{n}(0)|$ in the expression for 
$\bG^{<}$ we find
\be
-i G^<(\br,t;\br',t') = -i \sum_{m,n} \langle \br| \bSS(t;0) |\vf_m \rangle \langle \vf_m|
\bff^< | \vf_n \rangle \langle \vf_n| \bSS(0;t) \br'\rangle 
=
\sum_{n}f(\ve_{n})\vf_{n}(\br,t)\vf_{n}^{\ast}(\br',t'), \label{eq:g0}
\ee
which for $t=t'$ reduces to the time-dependent density matrix. The Green
function $\bG^>$ becomes
\be
-i G^>(\br,t;\br',t') = 
-\sum_{n}[1-f(\ve_{n})]\vf_{n}(\br,t)\vf_{n}^{\ast}(\br',t'), \label{eq:g0b}
\ee

Knowing the greater and lesser Green functions we can also 
calculate $\bG^{\ret,\adv}$. By definition we have 
$$
\bG^\ret (t;t')=\theta(t-t')
[\bG^{>}(t;t')-\bG^{<}(t;t')]
=-i \theta(t-t')\bSS(t,t'),
$$
and similarly
\begin{equation}
\bG^\adv (t;t')=i\theta(t'-t)\bSS(t,t')=
[\bG^\ret (t';t)]^{\dag}.
\label{gagr}
\end{equation}
In the above expressions the Fermi distribution function has 
disappeared. The information carried by $\bG^{\ret,\adv}$ is the same 
contained in the one-particle evolution operator. There is no 
information on how the system is prepared (how many particles, 
how they are distributed, etc). We use this observation to rewrite 
$\bG^{\lessgtr}$ in terms of $\bG^{\ret,\adv}$
\begin{equation}
\bG^{\lessgtr}(t;t')=\bG^\ret (t;0)\bG^{\lessgtr}(0;0)
\bG^\adv (0;t').
\label{r<a}
\end{equation}
Thus, $\bG^{\lessgtr}$ is completely known once we know how to 
propagate the one-electron orbitals in time and how they are 
populated before the system is perturbed \cite{blandin,cini,stefanucci}. 
We also 
observe that an analogous relation holds for $\bG^{\rceil,\lceil}$
$$
\bG^{\rceil}(t;\tau)=i\bG^\ret (t;0)\bG^{\rceil}(0;\tau),
\quad
\bG^{\lceil}(\tau;t)=-i \bG^{\lceil}(\tau;0)\bG^\adv (0;\tau).
$$

Let us now focus on a special kind of disturbance, 
namely $\bh(t)=\bh_{0}+\theta(t)\bh_{1}$. In this case
\begin{equation}
\bG^\ret (t;t')=-i \theta(t-t') e^{-i(\bh_{0}+\bh_{1})(t-t')}
\label{thp}
\end{equation}
depends only on the difference between the time arguments.
Let us define the Fourier transform of $\bG^{\ret,\adv}$ from
$$
\bG^{\ret,\adv}(t;t')=\int \frac{\D\omega}{2\pi} e^{-i\omega(t-t')}
\bG^{\ret,\adv}(\omega).
$$ 
The step function can be written as
$\theta(t-t')=\int\frac{\D\omega}{2\pi i}
\frac{e^{i \omega(t-t')}}{\omega-i\eta}$,
with $\eta$ an infinitesimally small positive constant. Substituting 
this representation of the $\theta$-function into (\ref{thp}) and 
shifting the $\omega$ variable one readily finds
$$
\bG^\ret (\omega)=\frac{1}{\omega-\bh_{0}-\bh_{1}+i\eta},
$$
and therefore $\bG^\ret (\omega)$ is analytic in the upper half 
plane. On the other hand, from (\ref{gagr}) it follows that 
$\bG^\adv(\omega)=[\bG^\ret(\omega)]^{\dag}$
is analytic in the lower half plane. What can we say about the 
greater and lesser component? Do they also depend only on the 
difference $t-t'$? The answer to the latter question is negative. 
Indeed, we recall that they contain information on how 
the system is prepared before $\bh_{1}$ is switched on. In 
particular the original eigenstates are eigenstates of $\bh_{0}$ 
and in general are not eigenstates of the Hamiltonian 
$\bh_{0}+\bh_{1}$ at positive times. From (\ref{r<a}) one can see that 
$\bG^{\lessgtr}(t;t')$ cannot be expressed only in terms of the time difference 
$t-t'$. For instance 
$$
\bG^{<}(t;t')=e^{-i(\bh_{0}+\bh_{1})t}i f(\bh_{0})
e^{i(\bh_{0}+\bh_{1})t'},
$$
and unless $\bh_{0}$ and $\bh_{1}$ commute, it is 
a function of $t$ and $t'$ separately.

It is sometimes useful to split $\bh(t)$ in two parts and treat one 
of them perturbatively. Let us think, for instance, of a system 
composed of two connected subsystems $A+B$. In case we know how 
to calculate the Green function of the isolated subsystems $A$ and $B$, 
it is convenient to treat the connecting part as a perturbation. 
Thus, we write $\bh(t)=\bcalE(t)+\bcalV(t)$, 
and we define $\bg$ as the Green function 
when $\bcalV=0$. The $\bg$ is a solution of 
$$
\left\{i \frac{\D}{\D z}-\bcalE(z)\right\}\bg(z;z')=
{\bf 1}\delta(z,z'),
$$
and of the corresponding adjoint equation of motion. Furthermore, the Green 
function $\bg$ obeys the KMS boundary conditions. With these 
we can use $\bg$ 
to convert the equations of motion for $\bG$ into integral 
equations
\be
\bG(z;z') = \bg(z;z')+\int_{\gamma}\D\bz\;\bg(z;\bz)\bcalV(\bz)\bG(\bz;z')
=
\bg(z;z')+\int_{\gamma}\D\bz\;\bG(z;\bz)\bcalV(\bz)\bg(\bz;z');
\label{dni}
\ee
the integral on $\bz$ is along the generalized Keldysh contour of 
Fig.~\ref{ke}. One 
can easily check that this $\bG$ satisfies both (\ref{lem}) 
and (\ref{rem}). $\bG$ also obeys the KMS boundary conditions 
since the integral equation is defined on the contour of Fig.~\ref{ke}.

In order to get some familiarity with the above perturbation scheme, 
we consider explicitly the system $A+B$ already mentioned. We partition the 
one-electron Hilbert space in states of the subsystem $A$ and states 
of the subsystem $B$. The ``unperturbed'' system is described by 
$\bcalE$, while the connecting part by $\bcalV$ and  
$$
\bcalE=\left[\begin{array}{cc}
\bcalE_{AA} & 0 \\
0 & \bcalE_{BB}
\end{array}\right],\quad\quad
\bcalV=\left[\begin{array}{cc}
0 &\bcalV_{AB}  \\
\bcalV_{BA} & 0
\end{array}\right].
$$
Taking into account that $\bg$ has no off-diagonal matrix 
elements, the Green function projected on one of the two subsystems, 
e.g., $\bG_{BB}$, is
$$
\bG_{BB}(z;z')=\bg_{BB}(z;z')+
\int_{\gamma}\D\bz\; \bg_{BB}(z;\bz)\bcalV_{BA}(\bz)\bG_{AB}(\bz;z')
$$
and
$$
\bG_{AB}(z;z')=
\int_{\gamma}\D\bz\; \bg_{AA}(z;\bz)\bcalV_{AB}(\bz)\bG_{BB}(\bz;z').
$$
Substituting this latter equation into the first one, we obtain a 
closed equation for $\bG_{BB}$:
\begin{equation}
\bG_{BB}(z;z')=\bg_{BB}(z;z')+
\int_{\gamma}\D\bz\D\bz' \bg_{BB}(z;\bz)\bS_{BB}(\bz;\bz')
\bG_{BB}(\bz';z'),
\label{gBB}
\end{equation}
with
$$
\bS_{BB}(\bz;\bz')=\bcalV_{BA}(\bz)\bg_{AA}(\bz;\bz')\bcalV_{AB}(\bz')
$$
the embedding self-energy. The retarded and advanced component can 
now be easily computed. With the help of Table \ref{kid} one finds
$$
\bG_{BB}^{\ret,\adv}=\bg_{BB}^{\ret,\adv}+
\bg^{\ret,\adv}_{BB}\cdot\bS^{\ret,\adv}_{BB}
\cdot\bG^{\ret,\adv}_{BB}.
$$

Next, we have to compute the lesser or greater component. 
As for the retarded and advanced components, this can be done 
starting from (\ref{gBB}). The reader 
can soon realize that the calculation is rather complicated, due to the 
mixing of pure real-time functions with function having one real time 
argument and one imaginary time argument, see Table \ref{kid}.  Below, we use
(\ref{r<a}) as a feasible short-cut. A closed equation for the retarded and
advanced component has been already obtained. Thus, we simply need an equation
for $\bG^{\lessgtr}(0;0)$. Let us focus on the lesser component 
$\bG^{<}(0;0)=i \bff^<$. Assuming that the Hamiltonian $\bh_0$ is
hermitian, the matrix $(\omega - \bh_0)^{-1}$ has poles at frequencies
equal to the eigenvalues of $\bh_0$. These poles are all on the real 
frequency axis, and we can therefore write
\begin{equation}
\bG^{<}(0;0)=i f(\bh_{0})=\int_\gamma \frac{d\zeta}{2\pi}
f(\zeta)  \frac{1}{\zeta-\bh_0},
\end{equation}
where the contour $\gamma$ encloses the real frequency axis.

\section{Action functional and TDDFT}

We define the  action functional
\be
\tilde A= i \ln \Tr \left\{e^{\beta \mu \hat N} \hat S (-i \beta;0)\right\}, 
\label{eq:action}
\ee
where the evolution operator $\hat S$ is the same as defined in (\ref{eo}).
The action functional is a tool for generating equations of motion, and is not
interesting \textit{per se}. Nevertheless, one should notice that the action,
as defined in (\ref{eq:action}) has a numerical value equal to $i \ln Z$, where
$Z$ is the thermodynamic partition function. 

It is easy to show that if we make a perturbation $\delta \hat V(z)$ in the
Hamiltonian, the change in the evolution operator is given by
\be
i \frac{\D}{\D z} \delta \hat S(z;z') = \delta \hat V(z) \hat S(z;z') + 
\hat H(z) \delta S(z;z').
\ee
A similar equation for the dependence on $z'$, and the boundary condition
$\delta \hat S(z;z) = 0$ gives
\be
\delta \hat S(z;z') = -i \int_{z'}^z d\bar z \, \hat S(z;\bar z) \delta \hat 
V(\bar z) \hat S(\bar z, z'). \label{eq:ds}
\ee
We stress that the time-coordinates are on a contour going from $0$ to          
$-i \beta$. The variation in, e.g., $V(t_+)$ is therefore independent of 
the variation in $V(t_-)$.
If we let $\delta \hat V(z) = \int d\br \, \delta  v(\br, z) \hat
n(\br)$, a combination of (\ref{eq:action}) and (\ref{eq:ds}) yields 
[compare to (\ref{ev7})] the expectation values of
the density,
\bea
\frac{\delta \tilde A}{\delta v(\br,z)}&=& 
\frac{i}{\Tr\left\{e^{\beta\mu \hat N} \hat S(-i \beta;0) \right\} }
\frac{\delta}{\delta v(\br,z)} 
\Tr\left\{e^{\beta\mu N} \hat S(-i \beta;0) \right\} \nonumber\\
&=& \frac{\Tr\left\{e^{\beta \mu \hat N} \hat S(-i \beta;0)
\hat S(0;z) \hat n(\br) \hat S(z,0) \right\}}
{\Tr\left\{e^{\beta\mu \hat N} \hat S(-i \beta;0) \right\} }
= n(\br,z).
\eea
A physical potential is the same on the positive and on the negative branch
of the contour, and the same is true for the corresponding 
time-dependent density, $n(\br, t)=n(\br, t_\pm)$. A
density response function defined for time-arguments on the contour is found
by taking the functional derivative of the density with respect to the
external potential. Using the compact notation $1=(\br_1, z_1)$, the response
function is written
\be
\chi(1; 2) = \frac{\delta n(1)}{\delta v(2)}
=\frac{\delta^2 \tilde A}{\delta v(1) \delta v(2)}= 
\chi(2; 1). \label{eq:chi}
\ee
This response function is symmetric in the space and time-contour
coordinates. We again stress that the variations in the potentials at $t_+$
and $t_-$ are independent. If, however, one uses this response function to
calculate the density response to an actual physical perturbing electric
field,  we obtain
\be
\delta n(\br,t)=\delta n(\br, t_\pm) = \int_\gamma dz' \int d\br' \chi(\br, t_\pm; \br' z') \delta v(\br', z'), \label{eq:resp1}
\ee
where  $\gamma$ indicates an integral along the contour.
In this expression, the perturbing potential (as well as the induced density
response) is independent of whether it is
located on the positive or negative branch, i.e. $\delta v(\br',t'_{\pm})=
\delta v(\br',t')$.  We consider a perturbation of a system initially in 
equilibrium, which means that $\delta v(\br',t')\neq 0 $ only for $t'>0$,
and we can therefore ignore the integral along the imaginary track of the
time-contour. The contour integral then consists of two parts: 1) First an integral
from $t'=0$ to $t'=t$, in which $\chi=\chi^>$, and 2) an integral from $t'=t$
to $t'=0$, where $\chi=\chi^<$. Writing out the contour integral in (\ref{eq:resp1})
explicitly then gives
\be
\delta n(\br,t) = \int_0^t dt' \int d\br' \left[
\chi^>(\br t; \br' t') -\chi^<(\br t; \br' t') \right]\delta v(\br',t')
= \int_0^\infty dt' \int d\br' \chi^\ret(\br t; \br' t') \delta v(\br',t').
\ee
The response to a perturbing field is therefore given by the retarded response
function, while $\chi(1,2)$ defined on the contour is symmetric in 
$(1 \leftrightarrow 2)$.

If we now consider a system of non-interacting electrons in some external 
potential $v_\KS$, we can similarly define a non-interacting action-functional
$\tilde A_\KS$. The steps above can be repeated to calculate the 
non-interacting response function. The derivation
is straightforward, and gives
\be
\chi_\KS(1; 2) =  \frac{\delta^2 \tilde A_\KS}{\delta v_\KS(1) \delta v_\KS(2)}
= -i G_\KS(1; 2) G_\KS(2;1).  \label{eq:chi0}
\ee
The non-interacting Green function $G_\KS$ has the form given in 
(\ref{eq:n-g}), (\ref{eq:g0}) and (\ref{eq:g0b}).
The retarded response-function is
\bea
\chi^\ret_\KS(\br_1,t_1 ; \br_2,t_2) 
 &=& -i \theta(t_1-t_2) \left[
G^>_\KS(\br_1, t_1;\br_2,t_2) G^<_\KS(\br_2, t_2;\br_1,t_1)  
 -G^<_\KS(\br_1, t_1;\br_2,t_2) G^>_\KS(\br_2, t_2;\br_1,t_1) \right]
\nonumber\\
&=&i \sum_{n,m} [f(\ve_m)-f(\ve_n)] \vf_n(\br_1,t_1) \vf^*_m(\br_1,t_1)
\vf_m(\br_2,t_2) \vf^*_n(\br_2,t_2),
\eea
where we have used (\ref{eq:g0}) and (\ref{eq:g0b}) in the last step.

Having defined the action functional for both the interacting and the 
non-interacting systems, we now make a Legendre transform, and define
\be
A[n]= -\tilde A[v]+\int d(1) \, n(1) v(1),
\ee
which has the property that $\delta A[n]/\delta n(1)=v(1)$. Similarly, we
define the action functional
\be
A_\KS[n]= -\tilde A_\KS[v_\KS]+\int d(1) \, n(1) v_\KS(1).
\ee
with the property  $\delta A_\KS[n]/\delta n(1)=v_\KS(1)$. 
The Legendre transforms assume the existence of a one-to-one correspondence 
between the density and the potential.
From these action functionals, we now define the exchange-correlation part to
be 
\be
A_\xc[n]=A_\KS[n]-A[n]-\frac{1}{2}\int d(12) \,
\delta(z_1,z_2)\frac{n(1) n(2)}{|\br_1-\br_2|}.
\ee
Taking the functional derivative with respect to the density gives
\be
v_\KS[n](1)=v(1)+v_\Hartree(1)+ v_\xc[n](1) \label{eq:vs}
\ee
where $v_H(1)$ is the Hartree potential and 
$v_\xc(1)=\delta A_\xc/\delta n(1)$. Again, for time-arguments on the real 
axis, these potentials are independent of whether the time is on the positive
or the negative branch. If we, however, want to calculate the response function
from the action functional, then it is indeed important which part of the
contour the time-arguments are located on.

We already described how to define response function on the contour, both
in the interacting (\ref{eq:chi}) and the non-interacting (\ref{eq:chi0}) case.
Given the exact Kohn-Sham potential, the TDDFT response function should give
exactly the same density change as the exact response function,
\be
\delta n(1) = \int d(2) \, \chi(1;2)     \delta v(2)
            = \int d(2) \, \chi_\KS(1;2) \delta v_\KS(2). \label{eq:resp}
\ee
The change in the Kohn-Sham potential is given by
\bea
\delta v_\KS(1) &=& \delta v(1) 
+ \int d(2)\, \frac{\delta v_\Hartree(1)}{\delta n(2)} \delta n(2)
+ \int d(2)\, \frac{\delta v_\xc(1)}{\delta n(2)} \delta n(2)
\nonumber\\
&=& \delta v(1) + 
\int d(2)\, f_\Hxc (1; 2) \delta n(2) 
= \delta v(1) + 
\int d(23)\, f_\Hxc (1; 2) \chi(2;3) \delta v(3) 
\eea
where $f_\Hxc(1;2)= \delta(z_1,z_2)/|\br_1-\br_2| +
\delta v_\xc(1)/\delta n(2)$. Inserted in 
(\ref{eq:resp}),  we obtain
\be
\chi(1;2) = \chi_\KS(1;2) + 
\int d(34) \, \chi_\KS(1;3) f_\Hxc(3;4) \chi(4;2).
\ee
This is the response function defined for time-arguments on the contour. 
If we want to calculate the response induced by a perturbing potential, the
density
change will be given by the retarded response function. Using Table \ref{kid},
we can just write down
\be
\chi^\ret (\br_1, t_1;\br_2,t_2) 
= \chi_\KS^\ret (\br_1, t_1; \br_2, t_2) 
+ \int dt_3 dt_4 d\br_3 d\br_4 \, \chi_\KS^\ret (\br_1, t_1;\br_3,t_3) 
f^\ret_\Hxc(\br_3, t_3;\br_4,t_4) \chi^\ret(\br_4, t_4;\br_2,t_2).
\ee
The time-integrals in the last expression go from $0$ to $\infty$. As
expected, only the retarded functions are involved in this expression.
We stress the important result that while the function  $f_\Hxc(1,2)$ 
is symmetric under the coordinate-permutation ($1\leftrightarrow 2$),
it is the retarded function
\bea
f^\ret_\Hxc(\br_1, t_1;\br_2, t_2) = 
\frac{\delta(t_1,t_2)}{|\br_1-\br_2|} + 
f_\xc^\ret(\br_1, t_1;\br_2,t_2)
\eea
which is used when calculating the response to
a perturbing potential.

\section{Example: Time-dependent OEP}

We will close this section by discussing the time-dependent optimized effective 
potential (TDOEP) method in the exchange-only approximation. This is a useful
example of how to use functions on the Keldysh contour. While the 
TDOEP equations can be derived from an action functional, we will here use the
time-dependent Sham-Schl\"uter equations as  starting point \cite{vanleeuwen96}.
This equation is derived  by employing a Kohn-Sham Green function, $G_\KS(1,2)$ 
which satisfies the equation of motion
\be
\left\{i \frac{\D}{\D z_1}+\frac{\nabla_1^2}{2}
-v_\KS(\br_1,z_1)\right\} G_\KS(\br_1, z_1;\br_2, z_2) 
=\delta(z_1,z_2)\delta(\br_1-\br_2), \label{eq:gks}
\ee
as well as the adjoint equation. The Kohn-Sham Green function is given
by (\ref{eq:g0}) and (\ref{eq:g0b}) in terms of the time-dependent 
Kohn-Sham orbitals.  Comparing (\ref{eq:gks}) to the Dyson-Schwinger
equation (\ref{lem}), we see that we can write an integral equation for
the interacting Green function in terms of the Kohn-Sham quantities,
\be
G(1;2) = G_\KS(1;2)+\int d(\bar 1 \bar 2)\, G_\KS(1;\bar 1) 
\left[\Sigma(\bar 1; \bar 2)
+\delta(\bar 1, \bar 2)[v(\bar 1)-v_\KS(\bar 1)] \right] G(\bar 2; 2).
\label{eq:intg}
\ee
It is important to keep in mind that this integral equation for $G(1;2)$ 
differs from the differential equations (\ref{lem}) and (\ref{rem}) in the
sense that we have imposed the boundary conditions of $G_\KS$ on $G$ in 
(\ref{eq:intg}). This means that if $G_\KS(1;2)$ satisfies the KMS boundary
conditions (\ref{bcl}), then so will $G(1;2)$.

If we now assume that for any density $n(1)=-iG(1;1^+)$ there is a potential
$v_\KS(1)$ such that $n(1)=-iG_\KS(1;1^+)$, we obtain the time-dependent 
Sham-Schl\"uter equation,
\be
\int d(\bar 1 \bar 2)\,  G_\KS(1;\bar 1) \Sigma(\bar 1; \bar 2)  G(\bar 2; 1)
= \int d(\bar 1) G_\KS(1;\bar 1) [v_\KS(\bar 1)-v(\bar 1)] G(\bar 1; 1).
\ee
This equation is formally correct, but not useful in practice since solving
it would involve first calculating the nonequilibrium Green function. Instead,
one sets $G=G_\KS$ and $\Sigma[G]=\Sigma[G_\KS]$. For a given self-energy functional,
we then have an integral equation for the Kohn-Sham equation. This equation
is known as the time-dependent OEP equation. Defining $\Sigma=
v_H+\Sigma_\xc$ and $v_\KS=v+v_\Hartree+v_\xc$, the TDOEP equation can
be written
\be
\int d(\bar 1 \bar 2)\,  G_\KS(1;\bar 1) \Sigma_\xc[G_\KS](\bar 1; \bar 2)  
G_\KS(\bar 2; 1)
= \int d(\bar 1) G_\KS(1;\bar 1) v_\xc(\bar 1) G_\KS(\bar 1; 1).
\label{eq:simpss}
\ee

In the simplest approximation, $\Sigma_\xc$ is given by the
exchange-only self-energy of Fig.~\ref{fexamples}a,
\be
\Sigma_\x(1;2)= i G^<_\KS(1;2)v(1,2)=- \sum_j
n_j \phi_j(1) \phi^*_j(2) w(1,2) 
\label{eq:exchange}
\ee
where $n_j$ is the occupation number. 
This approximation leads to what is known as the exchange-only TDOEP equations 
\cite{Ullrichetal:PRL95,Ullrichetal:BBG95,Gorling:PRA97}.
Since the exchange self-energy $\Sigma_\x$ is local in time, there  is  only one time-integration in (\ref{eq:simpss}). 
The x-only solution for the potential will be denoted $v_\x$.
With the notation
$\tilde \Sigma(3;4)=\Sigma_\x(\br_3 t_3;\br_4 t_3)-\delta(\br_3-\br_4) v_\x(\br_3 t_3)$
we obtain from (\ref{eq:simpss})
\bea
0 &=&  \int_{0}^{t_1} dt_3 \int d\br_3 d\br_4 \,  \left[ G_\KS^<(1;3)
\tilde\Sigma(3; 4) G_\KS^>(4;1)  \right.
- \left. G_\KS^>(1;3) \tilde\Sigma(3; 4) G_\KS^<(4;1) \right]  \nonumber\\
&& +\int_{0}^{-i \beta} dt_3 \int d\br_3 d\br_4 \,  G_\KS^\rceil(1;3) 
\tilde\Sigma(3; 4) G^\lceil_\KS(4;1) . 
\label{eq:oep1}
\eea
Let us first work out the last term which describes a time-integral from $0$
to $-i \beta$. On this part of the contour, the Kohn-Sham Hamiltonian is
time-independent, with $v_\x(\br,0)\equiv v_\x(\br)$, and 
$\vf_i(\br, t) = \vf_i(\br) \exp{(-i\ve_i t)}$. Since $\Sigma_\x$ is 
time-independent on this part of the contour, we can integrate
\be
\int_{0}^{-i \beta} dt_3 \,  G_\KS^\rceil(1;\br_3, t_3)
G_\KS^\lceil(\br_4, t_3;1) 
= -i \sum_{i,k} n_i (1-n_k) \vf_i(1) \vf^*_i(\br_3) \vf_k(\br_4) \vf^*_k(1)
\frac{e^{\beta(\ve_i-\ve_k)}-1}{\ve_i-\ve_k} 
\label{eq:result1}
\ee
If we then use  $n_i(1-n_k) (e^{\beta(\ve_i-\ve_k)}-1)=n_k - n_i$
and define the function $u_{\x,j}$ by
\be
u_{\x,j}(1)=-\frac{1}{\vf_j^*(1)} \sum_k n_k \int d2 \, \vf^*_j(2) \vf_k(2)
\vf^*_k(1) w(1,2)
\ee
we obtain from (\ref{eq:result1}) and (\ref{eq:exchange})
\be
\int_{0}^{-i \beta} dt_3 \int d\br_3 d\br_4 \,  G_\KS^\rceil(1,3)
\tilde \Sigma(3, 4) G_\KS^\lceil(4,1) 
=
 - \int  d\br_2 \,  \sum_j n_j \sum_{k\neq_j} 
\frac{\vf^*_j(\br_2) \vf_k(\br_2)}{\ve_j-\ve_k}
\vf_j(1) \vf_k^*(1)  \left[u_{\x,j} (\br_2) -v_{x}(\br_2) \right] 
+ c.c.
\ee
The integral along the real axis on the lhs of (\ref{eq:oep1}) can 
similarly be evaluated. Collecting our results we 
obtain the OEP equations on the same form as in \cite{Gorling:PRA97},
\bea
0&=&i \sum_j \sum_{k\neq j} n_j \int_{0}^{t_1} dt_2  \int d\br_2 \, 
\left[v_\x(2)-u_{\x,j}(2)\right] \vf_j(1) \vf_j^*(2) \vf_k^*(1) \vf_k(2) + c.c. \nonumber \\
&&+ \sum_j \sum_{k\neq j} n_j \frac{\vf_j(1) \vf_k^*(1)}{\ve_j-\ve_k} \int d\br_2 \,
\vf^*_j(\br_2) \left[v_\x(\br_2)-u_{\x,j}(\br_2)\right] 
\vf_k(\br_2) . \label{eq:oepfinal}
\eea
The last term represents the initial conditions, expressing that the system is
in thermal equilibrium at $t=0$. The equations have exactly the same form if
the initial condition is specified at some other initial time $t_0$. In the
case that we let $t_0 \to -\infty$, the term due to the initial conditions 
vanish and the remaining expression equals the one given in 
\cite{vanleeuwen96,Ullrichetal:PRL95,Ullrichetal:BBG95}.
The OEP-equations (\ref{eq:oepfinal}) 
in the so-called KLI-approximation have been
successfully used by Ullrich et al.\cite{Ullrichetal:BBG95} to 
calculate properties of atoms in strong laser fields.

\end{document}